\documentclass[preprint,12pt]{elsarticle}
\usepackage{tabularx}
\usepackage{graphicx}
\usepackage{color}
\usepackage[T1]{fontenc}
\usepackage{siunitx}
\usepackage{hyperref}
\hypersetup{
    colorlinks=true,
    linkcolor=blue,
    filecolor=magenta,      
    urlcolor=cyan,
    pdftitle={Overleaf Example},
    pdfpagemode=FullScreen,
    }
\usepackage[left=2.2cm, right=2.2cm]{geometry}
\usepackage{bm}
\usepackage{amsmath}
\usepackage{booktabs, multirow}

\newcommand{\bra}[1]{{\langle\, #1 \,\vert\,}}
\newcommand{\ket}[1]{{\,\vert\, #1 \,\rangle}}

\newcommand{\braHket}[3]{{\langle\, #1 \, \vert\, #2 \,\vert \, #3 \,\rangle}}

\usepackage{amssymb}

\journal{Computational Materials Science}

\begin{document}

\begin{frontmatter}

\title{Empirical Band-Gap Correction for LDA-Derived Atomic  Effective Pseudopotentials}


\author[inst1]{Surender Kumar}

\affiliation[inst1]{organization={Departments of Chemistry and Physics, Universität Hamburg},
            addressline={Luruper Chaussee 149}, 
            city={Hamburg},
            postcode={D-22761}, 
            country={Germany}}

\author[inst1]{Hanh Bui}

\author[inst1,inst2]{Gabriel Bester \corref{c1}}
\cortext[c1]{Corresponding author}
\ead{gabriel.bester@uni-hamburg.de}

\affiliation[inst2]{organization={The Hamburg Centre for Ultrafast Imaging},
            addressline={Luruper Chaussee 149}, 
            city={Hamburg},
            postcode={D-22761}, 
            country={Germany}}
\begin{abstract}
Atomic effective pseudopotentials enable atomistic calculations at the level of accuracy of density functional theory for semiconductor nanostructures with up to fifty thousand atoms. Since they are directly derived from \textit{ab-initio} calculations performed in the local density approximation (LDA), they inherit the typical underestimated band gaps and effective masses. We propose an empirical correction based on the modification of the non-local part of the pseudopotential and demonstrate good performance for bulk binary materials (InP, ZnS, HgTe, GaAs) and quantum dots (InP, CdSe, GaAs) with diameters ranging from 1.0 nm to 4.45 nm.  
Additionally, we provide a simple analytic expression to obtain accurate quasiparticle and optical band gaps for InP, CdSe, and GaAs QDs, from standard LDA calculation. 
\end{abstract}




\end{frontmatter}


\section{Introduction}

Atomic effective pseudopotentials (AEPs) \cite{cardenas12,karpulevich16,bester09} make it possible to perform atomistic calculations with the accuracy of density functional theory (DFT) for semiconductor nanostructures containing as many as fifty thousand atoms \cite{zirkelbach15}.
While this is achieved at the cost of a lower transferability (material specificity) and the lack of total energies and atomic forces (absence of self-consistent cycle), the approach delivers a high-quality electronic structure in the vicinity of the band gap (inner-eigenvalue solver). 
However, the use of the local density approximation (LDA) in the generation of the AEPs results in a significant underestimation of the band gaps and the effective masses of semiconductors and insulators. These errors are often substantial, with differences ranging up to 100$\%$ \cite{Martin04} from experimental values, and can even lead to the incorrect order of electronic states for HgTe  \cite{ sakuma11,fleszar05} and InAs \cite{zanolli07}.

Historically, great efforts have been made on the calculation of quasiparticle and optical gaps since the middle of the last century \cite{dyson49,hedin1965,salpeter51}. Based on many-body perturbation theory, the quasiparticle energies are described by the Dyson equation and are typically solved within the $GW$ approximation \cite{hedin1965,aryasetiawan1998,onida02,hybertsen86}. The optical gaps can be calculated \emph{ab initio}  by solving the Bethe-Salpeter equation (BSE) \cite{salpeter51,rohlfing00}, using time-dependent density functional theory (TDDFT) \cite{runge84} or the quantum Monte Carlo (QMC) method \cite{grossman01, williamson02}. Additional methods such as hybrid functionals \cite{becke88}, self-interaction correction (SIC) \cite{perdew81}, and LDA+U \cite{anisimov97} have also aimed at addressing the gap underestimation issue.  Although these approaches can give an accurate description of the fundamental and optical gaps, all of them are currently only possible for either bulk systems or molecules up to one hundred atoms, due to the high computational demand. 


As an alternative, various empirical corrections, retaining the computational simplicity of LDA, have been proposed. Such as corrections on the kinetic-energy density \cite{becke06, tran09,kim10}, correction to the local \cite{segev07,wang15},  and the non-local  \cite{wang1995,Fu1997,li05,Lany2008}  components of the different pseudopotentials.  
%
Motivated by the fact that the $GW$  results suggest that the LDA bandstructure is qualitatively correct up to a rigid energy shift of the conduction bands, the so-called scissor shift has been introduced \cite{baraff84, gunnarson86, johnson98}. While the underlying idea is simple, the operator (scissor-operator) fulfilling the task is non-trivial, and also non-local \cite{nastos05, godby88, bokdam16, bannow17, filip14}, making this approach computationally more demanding. Furthermore, a rigid shift will explicitly not correct the too-low effective masses, which is a significant drawback for nanostructures since confinement effects are directly linked to effective masses, as we will discuss further below.
\begin{figure}[h]
\centering
\includegraphics[width=0.49\columnwidth]{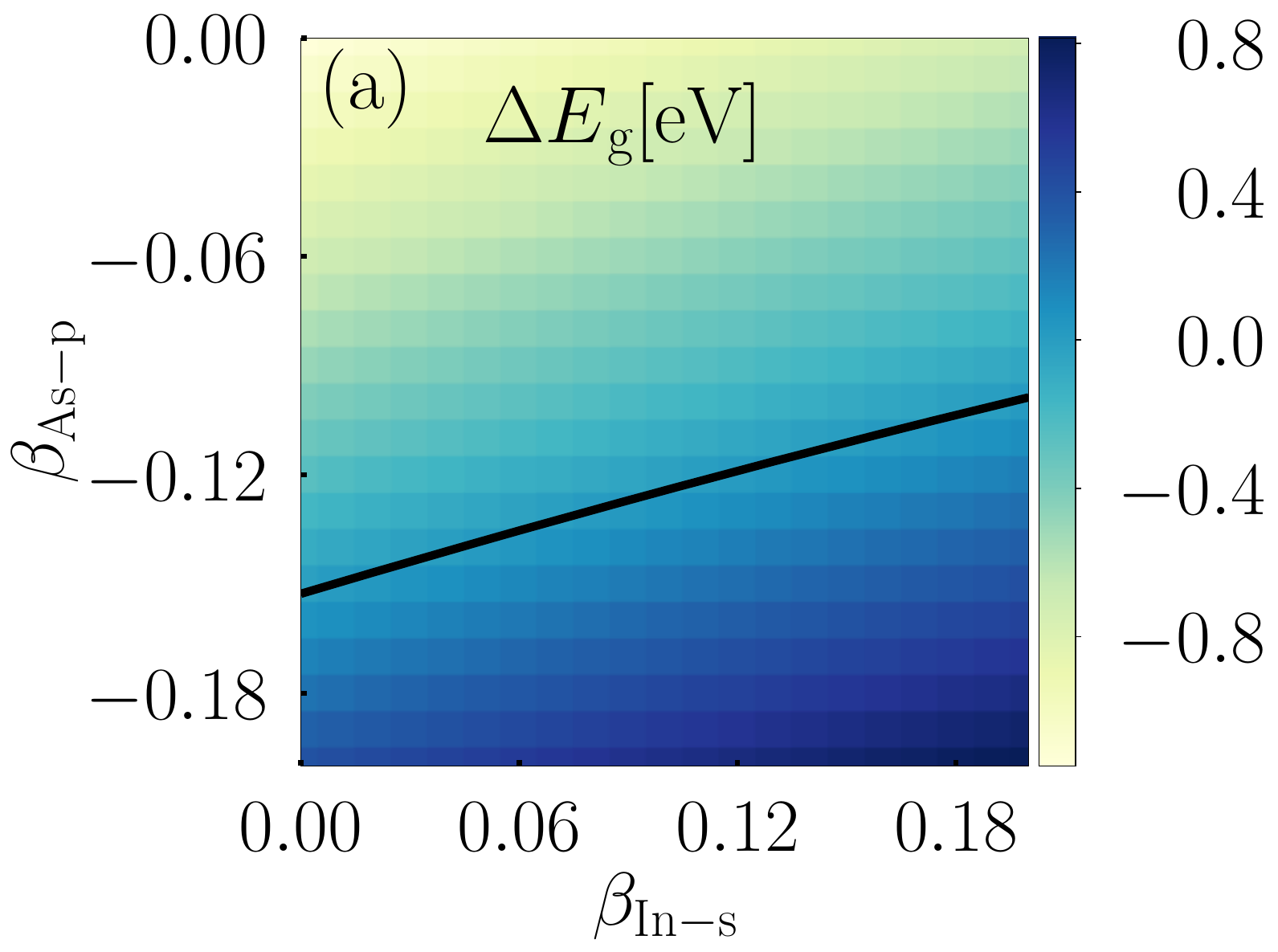}
\includegraphics[width=0.49\columnwidth]{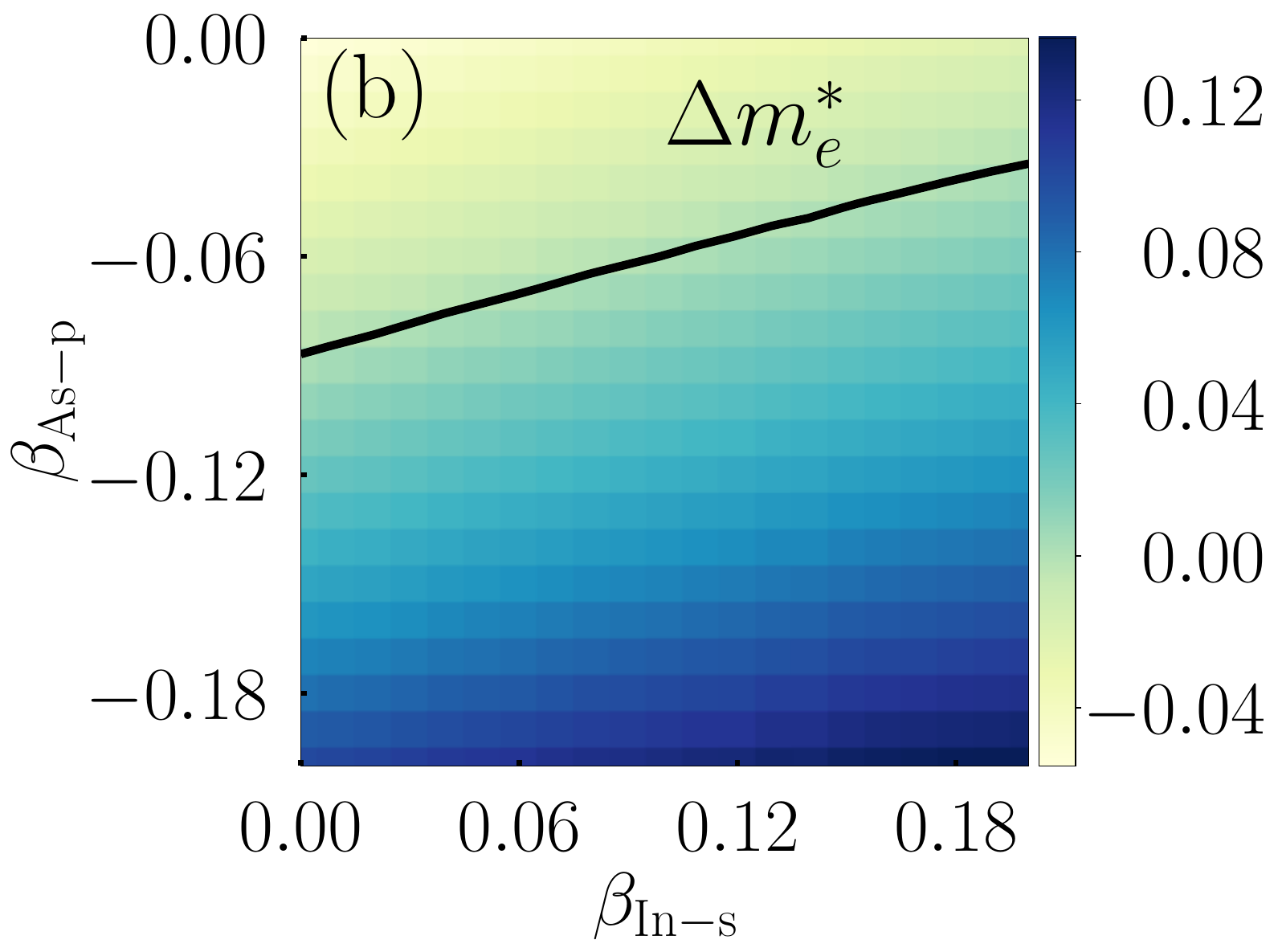}
\caption{ InP color-plot of the difference between the calculated and experimental values of (a)  the band-gap $\Delta  E_{ \mathrm{g}}$ and (b) the electron effective mass $\Delta  m^*_{ e}$ as a function of the $\beta$ parameters. The solid line corresponds to the zero-value contour line and represents the perfect match to the experiment.}
\label{fig:InP_gap}
\end{figure} 

In this work, we propose a simple empirical correction scheme to correct the band gap and improve the effective masses of LDA-derived AEPs. We validate our correction by a direct comparison to experiment for a) band energies at high-symmetry points in the Brillouin zone as well as the effective masses for bulk InP, ZnS, GaAs and HgTe and b) optical gaps and excitonic fine structure splitting  for InP, CdSe, and GaAs quantum dots with varying diameters.   
We use our findings to formulate a straightforward analytic expression, which can be adopted to correct the band gaps obtained using standard DFT codes in (small) quantum dots. Based on our excitonic screened configuration interaction (CI) results we further provide an expression to obtain accurate optical gaps. 

\section{Correction Scheme} 
\label{correction_scheme}

The AEPs are derived from the effective potential $\widehat{V}_{{\rm eff}}$ obtained as solution of the \textit{ab-initio} Kohn-Sham equation
\cite{kohn65}:
\begin{eqnarray}
 &  & \left(-\frac{\hbar^{2}}{2m}\Delta+\widehat{V}^{{\rm eff}}\right)\psi_{i}=\varepsilon_{i}\psi_{i},\nonumber \\
 &  & \widehat{V}^{{\rm eff}}=\widehat{V}^{{\rm ext}}+\widehat{V}^{{\rm Hartree}}[n]+\widehat{V}^{{\rm xc}}[n] \label{eq:effpot}
\end{eqnarray}
for a few different atomic configurations (see Refs.~\cite{cardenas12,zirkelbach15}). An advantage of the AEPs is that through the judicious choice of these few atomic configurations, an analytic connection exists between 
$\widehat{V}^\textrm{eff}(\bm G)$ and the AEP 
$\widehat{V}^\textrm{AEP}(|\bm G|)$ \cite{cardenas12}. 
Using the separable form of the norm-conserving pseudopotential formulated by Kleinman and Bylander \cite{hamann79,cardenas12,zirkelbach15,kleinman82}, we can express the effective self-consistent potential as:
\begin{align*}
\widehat{V}^\textrm{eff}  \approx 
\widehat{V}^\textrm{AEP} + 
\widehat{V}_{\rm NL} \quad ,
\end{align*}
with 
 \begin{equation}
\widehat{V}^\textrm{AEP} = \widehat{V}^{{\rm psp, loc}}+\widehat{V}^{{\rm Hartree}}+\widehat{V}^{{\rm xc}}
\end{equation}
and the non-local part: 
 \begin{equation}
\widehat{ V}_{\rm NL} = \sum_{l,m} \ket{l,m}\delta V_l(r) \bra{l,m} \quad ,
\end{equation}
where $\ket{l,m}$ are the spherical harmonics  and $\delta V_{l}$ is the
difference between the $l$-dependent pseudopotential  $V_{l}(r)$ and the
selected local part part already taken into account in $\widehat{V}^\textrm{AEP}$ (see Refs.~\cite{kleinman82,cardenas12}). 
This methodology has been shown to reproduce accurately the semi-local DFT results to within a few tens of meVs \cite{cardenas12,zirkelbach15,karpulevich16, bui20}.

In order to improve the band gap and the effective masses we decide to not modify $\widehat{V}^\textrm{AEP}$ but the non-local 
component of the pseudopotential according to:
\begin{equation}
    \delta V_l^\textrm{corr.} = 
    \begin{cases}
    \delta V_l(r) +\beta_l(1+\cos\frac{\pi r}{r_{\rm c}}) \quad &\mbox{for} \quad r<r_c \\
    0 & \mbox{for} \quad r\geq r_c \quad ,
    \end{cases}
\end{equation}
with $r_c$ = 2.25 Bohr and  $\beta_l$ as parameter. Our correction is therefore localized close to the atomic core where we expect less impact on the interatomic bonding. The idea to correct only close to the atomic core is in line with the atomic pseudopotential idea in general. Indeed, a match between the pseudopotential and the accurate all-electron result in the pseudopotential construction is only guaranteed beyond a cut-off radius similar to ours \cite{hamann79, troullier91}.   

In the majority of III-V and II-VI bulk semiconductors, the anion's p-orbital largely determines the valance band maximum (VBM), while the cation's s-orbital predominantly shapes the conduction band minimum (CBM). Our correction will therefore focus on the two parameters $\beta_ {\rm Cation-s}$ and $\beta_{\rm Anion-p}$.

The effect of the correction on the band gap and the electron effective mass is shown for InP in Fig.~\ref{fig:InP_gap}. 
The color code gives the deviation of the band gap (left) and the effective mass (right) from the experimental value ($E_{ \mathrm{g}}$=1.42 eV, $m^*_e$=0.082) as a function of the $\beta$-parameters. The thick line represents the contour line of zero error and shows a linear behavior in both cases of gap and electron-effective mass; a moderate value of the parameters allows us to obtain the correct masses and gaps. The figure also highlights a downside of the approach: both lines are nearly parallel pointing at the impossibility of correcting both masses and gaps with the same set of parameters. This behavior is also observed for the other materials investigated. While our correction is applied to the non-local part of the pseudopotential, it comes short of a truly non-local correction. The latter seems to be necessary to successfully correct both properties simultaneously, which would incur a notable increase in computational costs.
%

%

%
\begin{table*} [!hb]
  \centering
   \begin{tabular*}{\columnwidth}{l@{\extracolsep{\fill}}llllll}
    \hline
    \hline
     & & GaAs &InP &  ZnS   & ZnS  & HgTe\\
     & &  & &  (ZB) & (WZ) & \\
   
    \hline
    \hline
    \multirow{3}*{$E_\textrm{g}$ (eV) } & AEP & 0.359 &  0.270 & 1.760 & 1.930 & -1.300\\
       & AEP+$\beta$ & 1.520 & 1.420 & 3.720 &3.910 & -0.300\\
                                & Exp. & 1.520    {  \cite{madelung12}} & 1.420    {  \cite{madelung12}} & 3.720   {  \cite{madelung12}}&3.910    {  \cite{madelung12}}& -0.300   {  \cite{fleszar05,madelung12}}\\
                                 
 \hline
     & AEP & 0.730 &   1.180 & 3.280 & 3.880 & 1.500\\
    \multirow{1}{*}{$\Gamma_v-L_c$  }& AEP+$\beta$ &1.730 & 2.220 & 5.170 & 5.520&1.230 \\
    \multirow{1}*{(eV) }& Exp. & 1.850 & 1.930  &  & &\\
                                 & $GW$ &  &  & 5.010   {  \cite{klime14}}& &1.230   {  \cite{fleszar05}}\\
    \hline
     & AEP & 1.090 &   1.610 & 3.500 & &2.870\\
    \multirow{1}{*}{$\Gamma_v-X_c$ } & AEP+$\beta$ & 1.850 & 2.420 & 5.140 &&2.440\\
     \multirow{1}*{(eV) }                           & Exp. & 1.980  & 2.190  &  & & \\
                                 & $GW$ &   &  & 4.920   {  \cite{klime14}} & &2.450   {  \cite{fleszar05}} \\
    \hline
    
  \multirow{3}{*}{$\Delta_{\rm SO}$ (eV)} & AEP & 0.355 &   0.120 & 0.067 & 0.107&0.837\\
   & AEP+$\beta$ & 0.399 & 0.110 & 0.064 & 0.096& 0.856\\
                                & Exp. & 0.346  {  \cite{madelung12}} & 0.110   {  \cite{madelung12}}  &0.064   {  \cite{madelung12}} & 0.092   {  \cite{madelung12}}& 1.080   {  \cite{madelung12}}\\
                                     \hline
    \multirow{3}{*}{$\Delta_{\rm CF}$ (meV) } & AEP & & &  &27 &\\
    & AEP+$\beta$ &&  &  &25 &\\
                            & Exp. &  & &  &29   {  \cite{madelung12}}&\\
    \hline
    \multirow{3}{*}{$m_{\rm e}^{*} $ $ (m_{\rm 0})$ } & AEP & 0.023 &  0.026 & 0.140 & 0.140 & 0.216\\
   & AEP+$\beta$ & 0.096 & 0.130 & 0.340 & 0.320 & 0.024\\
   & Exp. & 0.066   {  \cite{kozhevnikov1995}} & 0.082   {  \cite{schneider1996}} & 0.220    {  \cite{imanaka1994}}&0.280  {  \cite{miklosz1967}} & 0.028   {  \cite{guldner1973}}\\
   
 \hline
    \multirow{3}{*}{$m_{\rm hh}^*$ $ (m_{\rm 0})$ } & AEP & 0.285 &  0.366  & 0.765 &1.596$^{a}$, 0.48$^{b}$ &  0.210\\
       & AEP+$\beta$ &0.362 & 0.473&1.396 &1.990$^{a}$, 0.61$^{b}$ & 0.337\\
    & Exp. & 0.340    { \cite{shanabrook1989}} & 0.450    {  \cite{rochon1975}}& 1.760    {  \cite{lawaetz1971}} &1.400$^{a}$, 0.49$^{b}$   {  \cite{miklosz1967}} &0.320   {  \cite{groves1967}}\\
    \hline
    
    \multirow{3}{*}{$E_{\rm g}/m_{\rm e}^{*}$} & AEP & 15.608 & 10.384 & 12.571 & 13.785&6.018\\
                               & AEP+$\beta$ & 15.833 & 10.923 & 10.941 &12.218& 12.500\\
                                & Exp. & 23.030  & 17.317  & 16.909 & 13.964&10.714\\
\hline
\hline
\end{tabular*}
 \caption{Calculated band gaps and effective masses without correction (AEP)  and with correction (AEP+$\beta$) for different high symmetry points compared with experimental and $GW$ results. For ZnS WZ, the superscripts $a$ and $b$ for the hole effective masses indicate the reciprocal space direction [001] and [010] respectively.}
\label{tab:egaps}
\end{table*} 
In this work, we choose to optimize the band gap at the cost of having somewhat too large effective masses. The opposite procedure, to optimize the effective masses at the cost of having too low band gaps is a viable alternative that may be advantageous if the extracted physical observable depends strongly on the masses and less on the gaps. 

Figure~\ref{fig:InP_gap}a) shows that any pair of correction parameters ($\beta_ {\rm In-s}$, $\beta_{\rm As-p}$) residing on the solid line yields the exact experimental band gap. We now face the question related to the appropriate selection of this pair. Since $\beta_ {\rm In-s}$ ($\beta_{\rm As-p}$) is almost directly proportional to the CBM (VBM) shift, we use $GW$ corrections to establish the appropriate weight of the CBM/VBM corrections, but fit the band gap to experiment and not to $GW$. We proceed as follows.

 We initiate the selection by quantifying the individual shifts in the CBM and  VBM between LDA and $GW$, which can be determined either through direct computation or by referencing pertinent literature \cite{klime14, sakuma11, fleszar05}. Subsequently, we proceed to compute the relative contributions (in $\%$) of the VBM and CBM shifts to the overall band gap difference between LDA and $GW$. Upon determining the relative shifts, we leverage these values in the correction procedure to correct the VBM and CBM energy levels in AEPs to fit with the experimental gap. 
For example, consider GaAs, which shows a 0.87 eV gap difference between the LDA gap  (0.32 eV)  \cite{klime14}  and $GW_0$ gap (1.19 eV) \cite{klime14} (experimental $E_{ \mathrm{g}}$ = 1.52 eV). The VBM shift of -0.78 eV in $GW_0$ w.r.t. LDA contributes 90$\%$ to the total gap difference, while the remaining 10$\%$ is contributed by the CBM. The $\beta_l$-parameters are chosen accordingly: 90$\%$ VBM shift and 10$\%$ CBM shift to obtain the experimental gap (and not the $GW$ gap, which can be off by several tens of percent, in this example 21\%). 
\begin{figure} [h]
\centering
\includegraphics[width=0.45\linewidth]{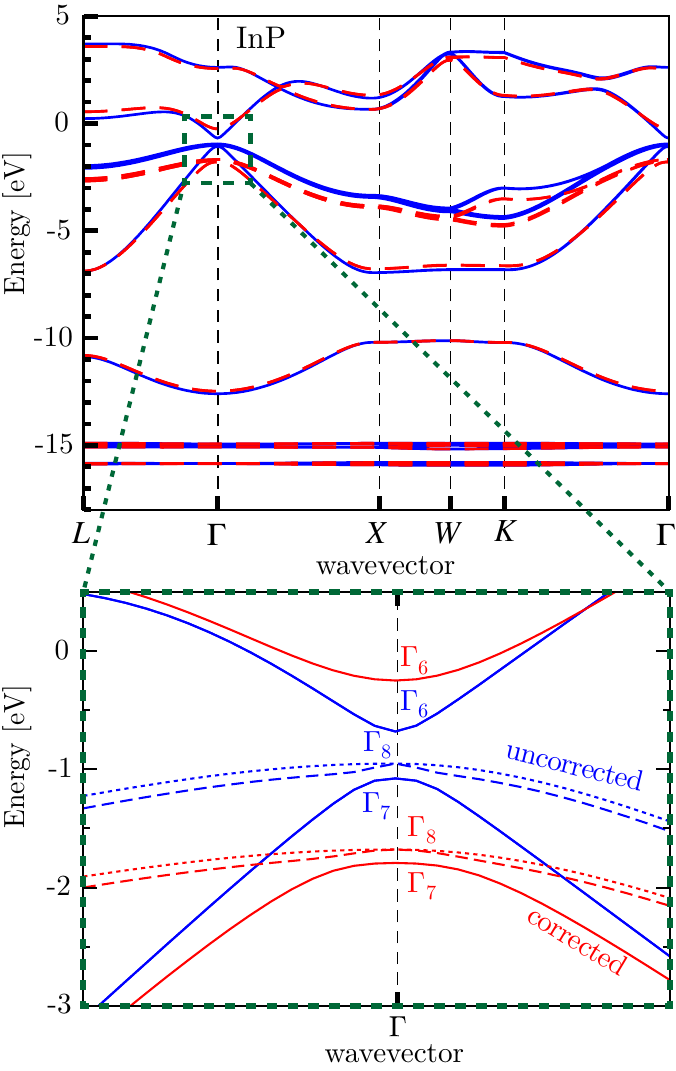}
\includegraphics[width=0.425\linewidth]{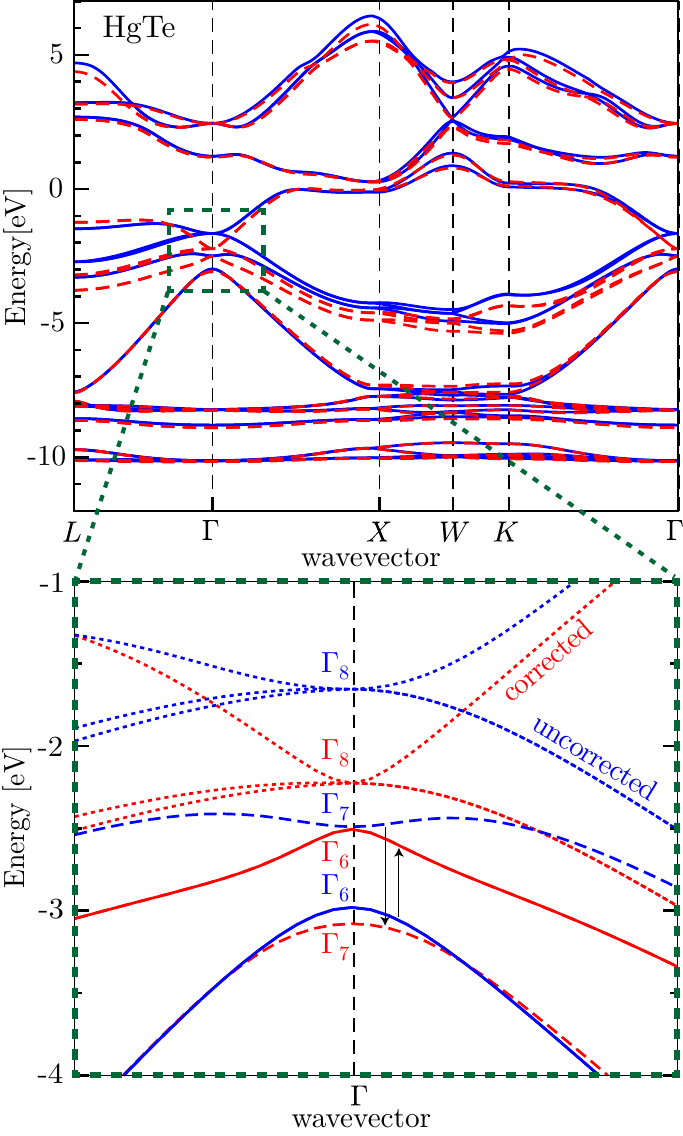}
\caption[Band structure of InP before and after corrections.]
{Band structure of zinc-blende InP (left)  and HgTe (right) calculated before (blue line) and after the $\beta$-correction (red dashed line). Bottom: magnification of the area in the green square. }
\label{fig:InP_bulk}
\end{figure} 
\section{Effect of the correction on the band structure}

To illustrate the impact of the correction on the band structure we show in  Fig.~\ref{fig:InP_bulk}  the bandstructures of InP and of the semi-metal HgTe.
The comparison between the uncorrected (blue) and the corrected (red) bands shows an almost rigid shift of the top of the valence bands to lower energies across the entire Brillouin zone. Further bands are significantly less affected. 
A qualitative alteration of the bands is primarily seen close to the band gap. 
For InP (Fig.~\ref{fig:InP_bulk}) the CBM is shifted up in energy and the curvature of the band is reduced (effective mass increased). 
For HgTe (Fig.~\ref{fig:InP_bulk}) we see very significant energy shifts and a reordering of the bands around the $\Gamma$-point. The DFT calculations, and hence our AEPs, predict a band inversion of the $\Gamma_{6}$ and $\Gamma_{7}$ bands, resulting in a qualitatively incorrect order of states. The $\beta$-correction successfully tackles this issue by restoring the accurate order of the $\Gamma_{6}$ and $\Gamma_{7}$ bands by shifting the $\Gamma_6$ band up and the $\Gamma_7$ band down in energy (following the arrows in Fig.~\ref{fig:InP_bulk}).

We have summarized further important band structure properties including band gaps at different symmetry points, spin-orbit splitting ($\Delta_{\rm SO}$), crystal field splitting for wurtzite (WZ) structures ($\Delta_{\rm CF}$), and effective masses for GaAs, InP, ZnS, and HgTe in Table  \ref{tab:egaps}. 
The table shows that our correction improves the $\Gamma_v-L_c$ (valence band top at $\Gamma$ to bottom of the conduction band at $L$) and the $\Gamma_v-X_c$ gaps significantly, reducing the error from approximately  40~\% to about 5~\%.
The pseudopotentials for spin-orbit interaction are directly taken from DFT-derived norm-conserving pseudopotentials and not modified in our AEP methodology \cite{zirkelbach15}. The agreement of the spin-orbit splitting $\Delta_{\rm SO}$ with the experiment is generally very good. 
\begin{figure}[h]
\centering
\includegraphics[width=0.70\linewidth]{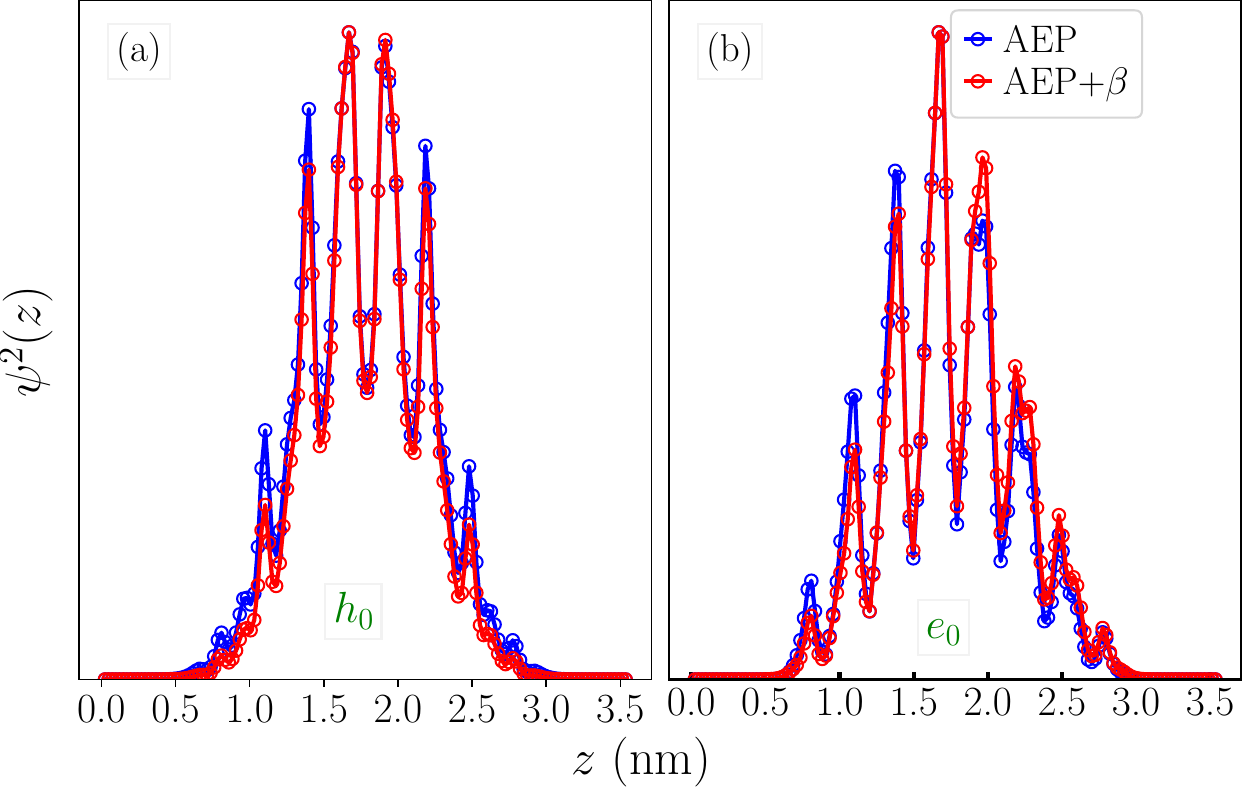}
\caption{Square of the wavefunction plotted along the [001]-direction for (a)  the  $e_0$ state and (b) the $h_0$ state of an InP QD with diameter 1.97 nm.}
\label{fig:InP_QD_wf}
\end{figure} 

The electron effective masses tend to be too large after the correction. While they are 36-68 \% too small at the DFT level, they are 14-58 \% too large after the correction. The hole-effective masses are generally improved by the correction with the exception of ZnS WZ. The results for HgTe have to be assessed separately, since the correction changes the order of the bands. In this case, both electron and hole effective masses are in good agreement with the experiment.

\section{Effect of the correction on the wavefunctions}
The modification of the non-local part of the pseudopotential, and hence the Hamiltonian, leads to new eigenfunctions. While it is difficult to assess if the new wavefunctions represent an improvement (this would require high quality self-consistent $GW$ calculations), we demonstrate in the following the extent of the modification by looking at wavefunction overlap and Coulomb matrix elements in quantum dots. 

\subsection{Bulk}

The overlap between the uncorrected and the corrected wavefunction is above 99\% for $e_0$ and $h_0$ for all the materials (see Supporting Information \cite{noteSI}).
The fact that the wavefunctions are only marginally modified by the correction can be assessed by the ratio  
$E_\textrm{g}/m_{e}^{*}$, shown at the bottom of Tab.~\ref{tab:egaps}, which is nearly unaffected by the correction (HgTe being a special case). Indeed, according to 
$k \cdot p$ perturbation theory \cite{cardona10}:
\begin{equation}
      \frac{m_0}{m^*_e} \approx \frac{2}{m_0 }\frac{P_{cv}^2}{E_{gap}}, \label{eq:kp}
\end{equation}
with $P_{cv} = \braHket{u_c}{\hat{p}}{u_v}$ and the Bloch function $u_{c,v}$ for the conduction and the valence bands, letting us expect a constant ratio of $E_\textrm{g}/m_{e}^{*}$. 
Note that the deviation of approximately 50$\%$  observed for HgTe can be attributed to the incorrect ordering of the $\Gamma_6$ and $\Gamma_7$ states in the AEP (LDA), as presented in Fig.~\ref{fig:InP_bulk}.

%
\subsection{Quantum Dots} 
\label{quantum_dots}
The AEP methodology is aimed at the study of zero-dimensional nanostructures which typically have a large number of atoms and cannot be addressed at the LDA level. We therefore investigate the influence of the correction on QD properties and especially how the correction influences the wavefunctions. We computed the single-particle energies and wavefunctions using the LATEPP package \cite{zirkelbach15} in combination with the AEP approach \cite{cardenas12, karpulevich16}. All calculations were performed with unrelaxed geometries with a minimum separation of \SI{6}{\angstrom} between periodically repeated quantum dots. To passivate the dangling bonds of the QDs, we employed fractional charge non-spherical pseudo-hydrogen (see Ref.~\citenum{karpulevich16}). All the QDs have a zinc-blende (ZB) crystal structure.
 \begin{table} [h]
    \centering
    \begin{tabular}{c c c c}
    \hline
    \hline
      Diameter (nm)   & $J_{e_0h_0}$ (AEP) & $J_{e_0h_0}$ (AEP+$\beta$) & Diff. in \%  \\
      \hline
      \hline
      1.61 &0.324 & 0.328 & 1.2 \\
      \hline
      1.97 & 0.251   & 0.257 & 2.3 \\
      \hline
      2.91 & 0.150   & 0.154 & 2.7\\
      \hline
      3.18 &  0.142   &  0.146 & 2.8\\
      \hline
      3.74 & 0.116  & 0.123 & 6.0 \\
      \hline
      4.45 &  0.094  & 0.102 & 8.5\\
      \hline
      \hline
    \end{tabular}
    \caption{Coulomb integral $J_{e_0h_0}$ (in eV) between the $e_0$ and the $h_0$ states calculated before and after the corrections for InP QDs with varying diameters.} 
    \label{tab:jval}
\end{table}
In Fig.~\ref{fig:InP_QD_wf}, we show the $e_0$ (lowest unoccupied QD orbital) and $h_0$ (highest occupied QD orbital) wavefunctions for an InP QD with 1.97 nm diameter before (blue) and after the band gap correction (red). We see relatively small changes, with a tendency for the corrected wavefunctions to be more localized, in agreement with their larger effective masses. A simple calculation of the wavefunction overlap between corrected and uncorrected wavefunctions shows deviations from the unity of less than 2\% (see supporting information \cite{noteSI}).

To understand how these alterations in the wavefunctions impact the energy contributions, we carried out calculations of the Coulomb integral ($J$)  between the electron ($e$) and hole ($h$) states, a parameter intrinsically dependent on the wavefunctions \cite{bester09,franceschetti99}, as defined in Eq.~(\ref{J_eq}) :
\begin{equation}
J_{he,h^{'}e^{'}} = e^{2}  \sum_{\sigma_1,\sigma_2}  \iint 
\frac{\psi^{*}_{h^{'}}(\mathbf{r_1}, \sigma_1) \psi^{*}_{e}(\mathbf{r_2}, \sigma_2) \psi_{h}(\mathbf{r_1}, \sigma_1) \psi_{e^{'}}(\mathbf{r_2}, \sigma_2)}{\varepsilon(\mathbf{r_1}, \mathbf{r_2})|\mathbf{r_1} - \mathbf{r_2}|} 
   d\mathbf{r_1} d\mathbf{r_2},
\label{J_eq}
\end{equation}
where $\sigma_1$, $\sigma_2$ are spin indices and  $\varepsilon(\mathbf{r_1}, \mathbf{r_2})$  is the microscopic screening function accounted for via the modified Penn-Resta-Haken approach \cite{bui20, franceschetti99}.  For a consistent comparison, $J$ values with and without $\beta$-correction were calculated with the same screening function. 
As shown in Table \ref{tab:jval} the Coulomb integrals increase slightly when the correction is applied, which is consistent with the associated increase in the effective mass and the stronger localization of the carrier.

%
%
\section{Comparison to experiments} \label{com_exp}

\subsection{Optical Band Gap}

\begin{figure}[!hb]
\centering
\includegraphics[width=0.70\linewidth]{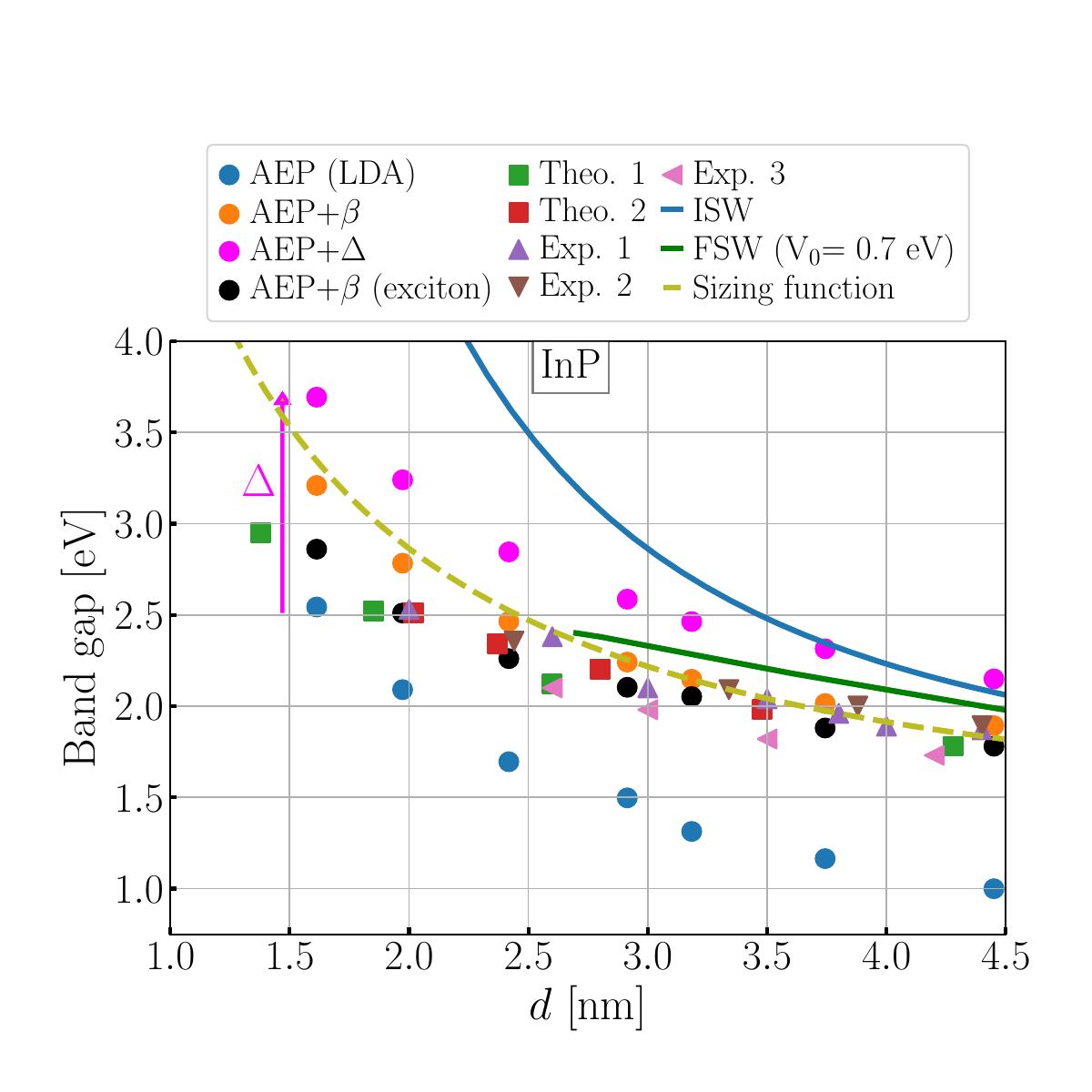}
\caption{Optical band gap of ZB InP QDs obtained from calculations at different levels of theory (see text for details, the black filled circles being the final theoretical result), compared with earlier theoretical work (Theo. 1   \cite{fu98}, Theo. 2   \cite{fu97}),  with the ``sizing function'' from Ref.~\citenum{aubert22} and with experimental work (Exp. 1 \cite{fu97}, Exp. 2 \cite{cho13} and Exp. 3 \cite{micic97}). ISW and FSW are effective mass results for infinite and finite spherical wells, respectively.}
\label{fig:InP_QD} 
\end{figure}

In Fig.~\ref{fig:InP_QD}, we present a comparative analysis of our results, obtained both with and without our $\beta$-correction,
with available experimental, theoretical, and spherical well approximation (both infinite and finite) literature values.  
\vfill
Our final exciton results, including the $\beta$-correction to the band gap and correlation effects using the screened CI approach (black-filled circles), agree very well with the experiments (triangles), as well as with other theoretical calculations based on the EPM approach \cite{fu98, fu97} (squares).
The quasiparticle results (orange, band-gap corrected single particle results) overshoot the optical band gap, as expected, due to the lack of Coulomb $e-h$ binding, while the LDA results (blue filled circles) yield too small gaps, as a well known consequence of the semi-local LDA approximation. 
In magenta, we show the results using the ``scissor'' shift, meaning in our case a simple rigid energy shift of the conduction band states with respect to the valence band states by an energy $\Delta$. The energy $\Delta$ is the difference between the bulk experimental band gap and the LDA (or AEP) band gap. This is a commonly used procedure to adjust the band gaps obtained at the LDA or GGA level. We observed that the scissor shift method tends to strongly overestimate the band gap of small QDs. 
The results from the infinite square well (ISW) model (solid blue line) significantly overestimate the band gap, which is generally known. On the other hand, the finite square well (FSW) model (solid green line) can be made to approximately fit the experimental results by using a well depth of 0.7 eV. This latter value is significantly lower than the correct value of several eV, which underlines the limitations of the continuum model description.

\begin{figure} [h]
\centering
\includegraphics[width=0.72\linewidth]{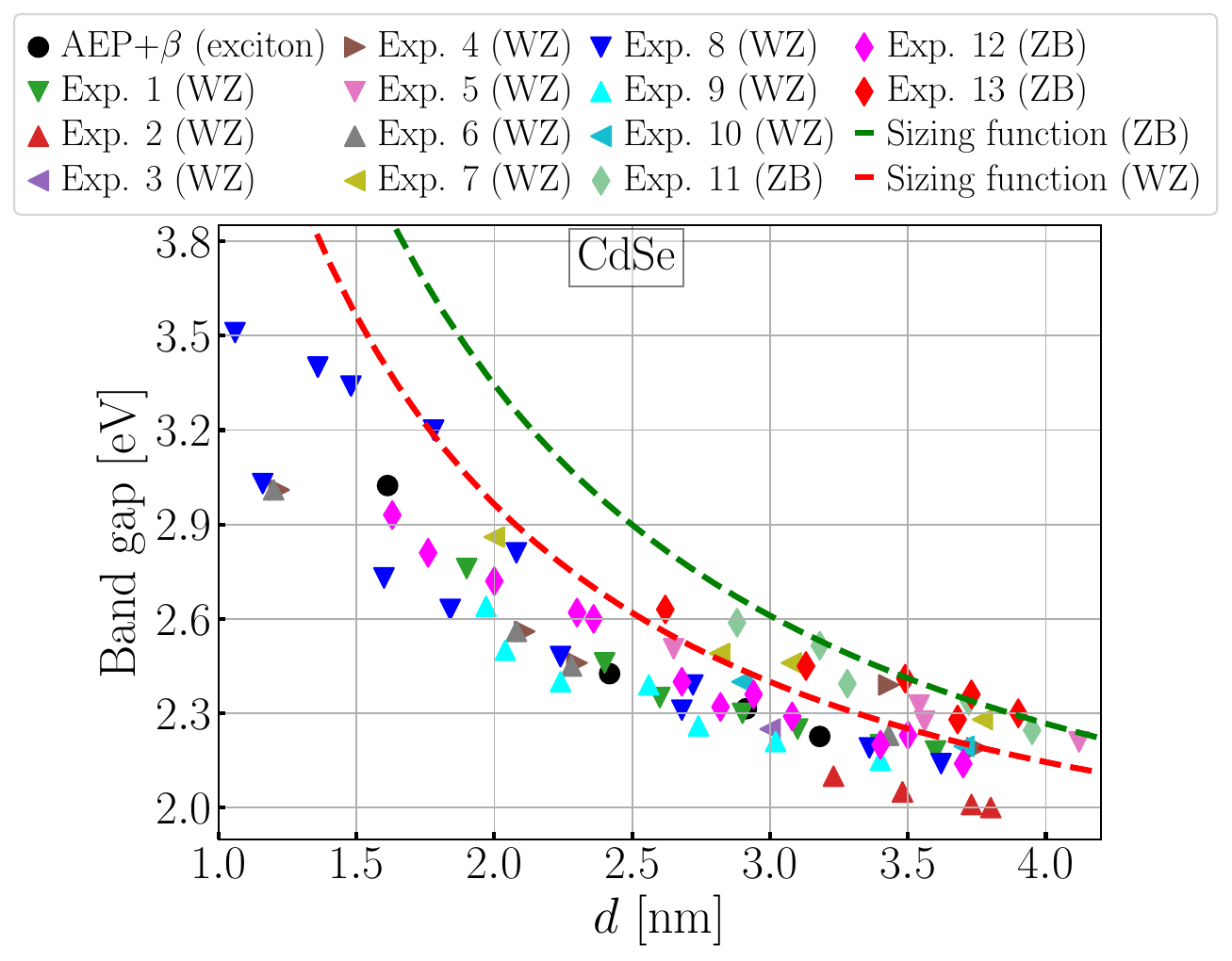}
\caption{Akin Fig.~\ref{fig:InP_QD} but for ZB CdSe QDs.  The references corresponding to the experiment number in the figure label are as follows:  1~\cite{inamdar08},  2 ~\cite{kucur2003},  3 ~\cite{querner05},  4~\cite{murray93},  5~\cite{aubert22},  6~\cite{karpulevich19},   7~\cite{meulenberg09},  8 ~\cite{yu03},  9 ~\cite{jasieniak09},  10~\cite{ning11},  11 ~\cite{aubert22}, 12 ~\cite{capek10}, and  13 ~\cite{aubert22}. Sizing function:  \cite{aubert22}.}
\label{fig:cdse_gaps}
\end{figure}

In Fig.~\ref{fig:cdse_gaps}, we show the corresponding results for CdSe QDs using a similar nomenclature. The QDs can have either ZB or WZ structure while our theoretical results are for ZB structures.  We notice a relatively large spread of the experimental results in general, which can be attributed, e.g., to the intrinsic difficulty to asses the QD size, the crystal structure, and the organic capping environment. Our theoretical results are found in the middle of the experimental data. The ``sizing function'' \cite{aubert22} for ZB yields larger band gap values for this size range (while it fits the results more accurately for larger QD sizes). 

\subsection{Splitting of the lowest two-electron states (``S-P''-splitting)}

\begin{figure} [h]
\centering
\includegraphics[width=0.63\linewidth]{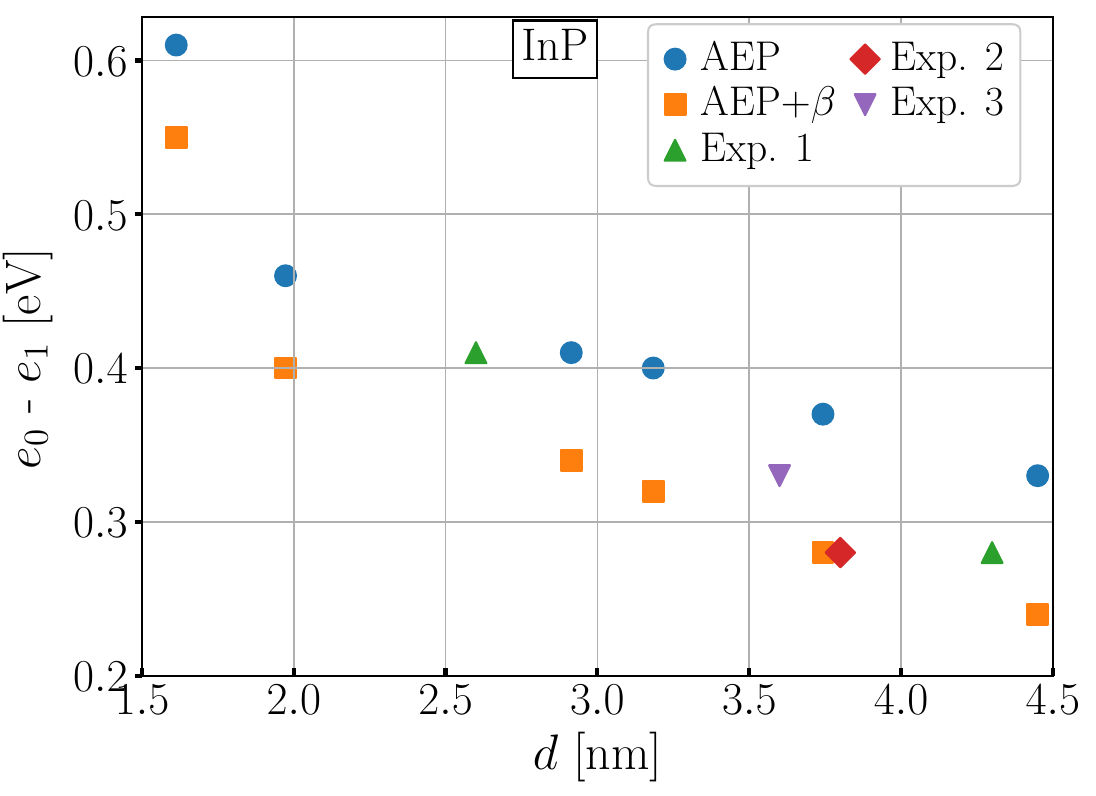}
\caption{Comparison of the energy splitting between the $e_0$ and  $e_1$ states for InP QDs as a function of diameter $d$. Experimental results Exp. 1 and Exp. 2 are from \cite{rumbles01} and \cite{blackburn05}, respectively.}
\label{fig:InP_QD_SP}
\end{figure}

Our simple $\beta$-correction leads to an accurate band gap description and we anticipate an improved description of the intraband energy splittings (compared to LDA) as well. 
In Fig.~\ref{fig:InP_QD_SP}, we present the splitting of the lowest two 
 unoccupied ($e_0$  and $e_1$) single-particle states (S-P splitting) for InP QDs calculated before (blue) and after (orange) $\beta$-correction along with the experimental results (red, green, violet).

The uncorrected AEP (and LDA) results lead to overestimated S-P splittings. This is also expected since the intraband S-P splitting is approximately inversely proportional to the effective mass, which is significantly too small at the LDA level. 
The corrected results (orange squares) are significantly lower and in better agreement with the experimental data. As we noticed earlier, when the band gap is corrected the effective mass tends to overshoot (become larger than the experimental value), hence our S-P splitting tends to be lower than the experimental value.
\subsection{Fine structure splitting}
 
The fine structure splitting (FSS) describes the small (but important) splitting of the ``ground state" exciton and is due to the $e$-$h$ exchange interaction in the presence of spin-orbit coupling and is an atomistic effect \cite{franceschetti99, bester03,bui20, torben23}. As anticipated for high-quality ZB InP QDs with spherical shape and $T_d$-symmetry   \cite{franceschetti99, bui20}, our calculations yield a 5-fold spin-forbidden dark state and a 3-fold spin-allowed bright state both with perfect degeneracy, strongly indicating that our correction scheme preserves symmetry. 

\begin{figure}[h]
\centering
\includegraphics[width=0.62\linewidth]{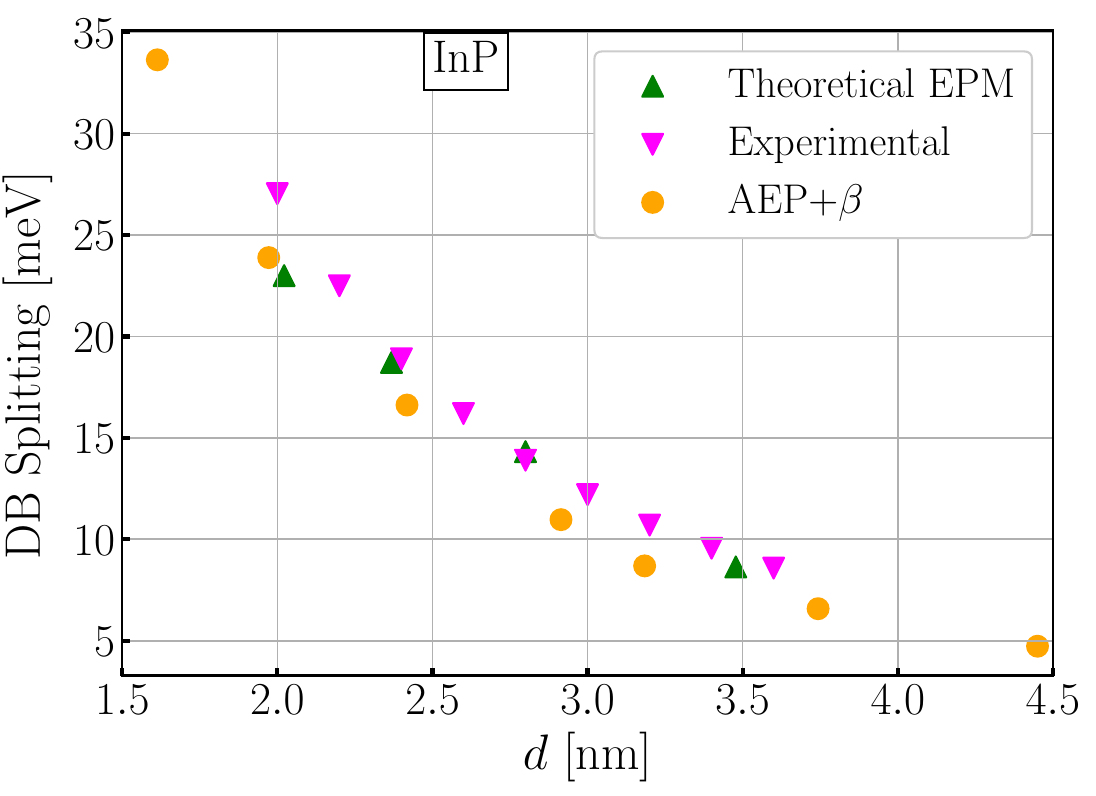}
\caption{Calculated dark-bright (DB) splitting of the lowest excitonic states of InP QDs as a function of diameter. Experimental \cite{micic97}  and theoretical EPM \cite{franceschetti99} results are shown in magenta and green, respectively.}
\label{fig:InP_QD_FS}
\end{figure}

 O. Mićić \textit{et al.}~\cite{micic97} have successfully synthesized high-quality, defect-free InP quantum dots
 that are exceptionally well-suited for a direct comparison. This early experimental work is exceptional, since ligands and atomic details of the quantum dot surface can significantly affect the FSS \cite{torben23}.  Theoretically, Franceschetti \textit{et al.}~\cite{franceschetti99} have computed the FSS of InP quantum dots assuming an ideal surface passivation (pseudohydrogens) and used a high-quality atomistic EPM method. In Fig.~\ref{fig:InP_QD_FS}, we show both results along with our calculations and see a general very good agreement.

\section{Suggested empirical correction to DFT (LDA/GGA) results} 
\label{sec:LDA_correction}

\subsection{Quasiparticle Band Gap}

\begin{figure*} [h]
\centering
\includegraphics[width=0.32\linewidth]{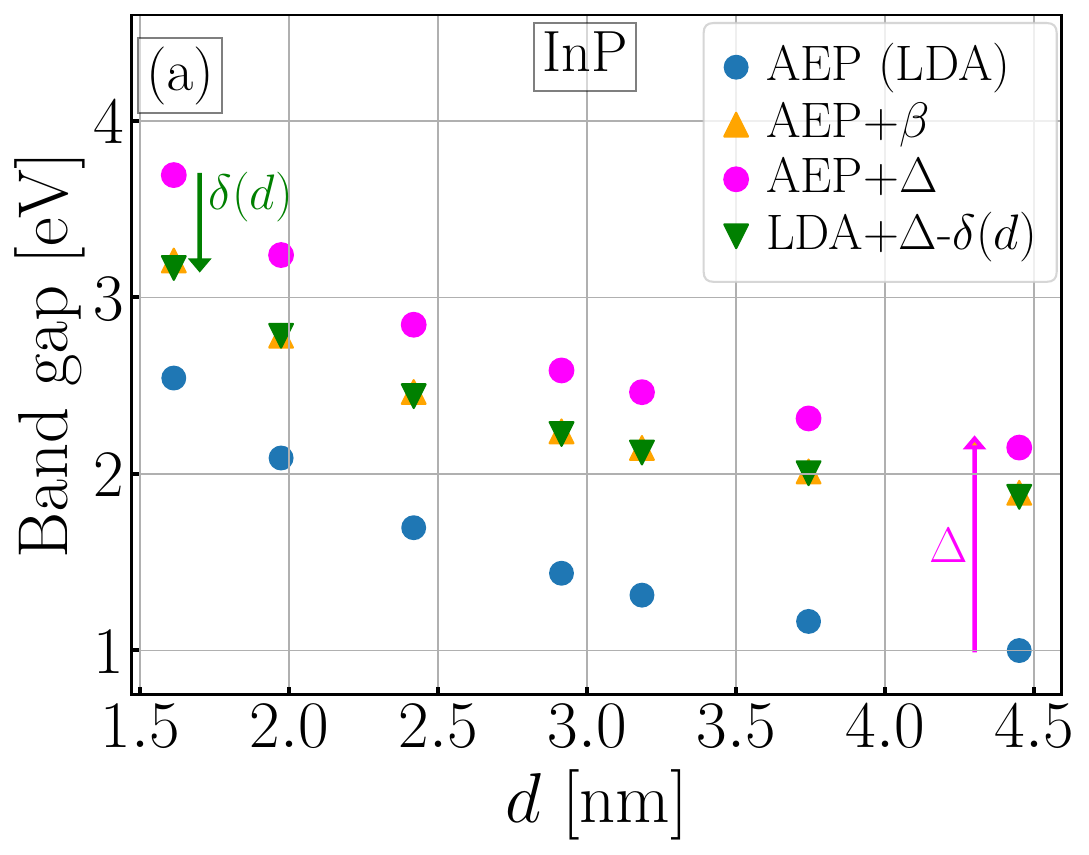}
\includegraphics[width=0.32\linewidth]{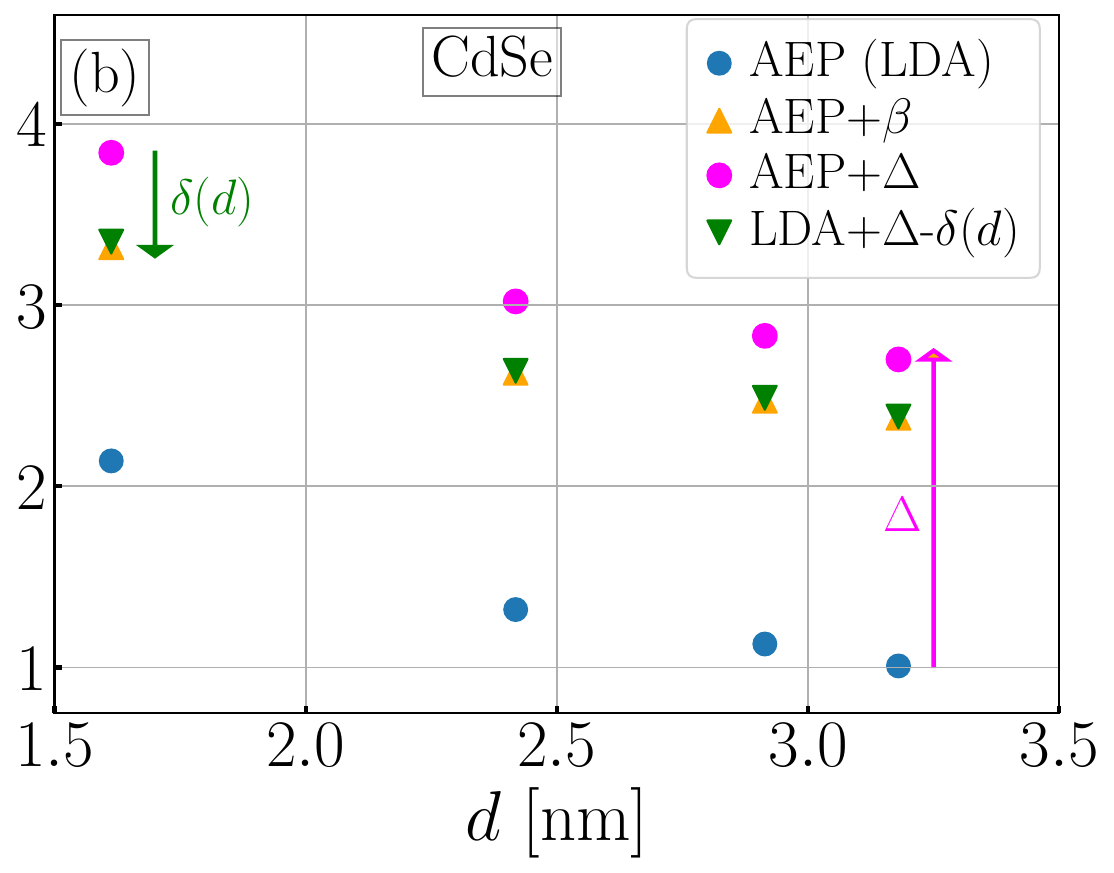}
\includegraphics[width=0.32\linewidth]{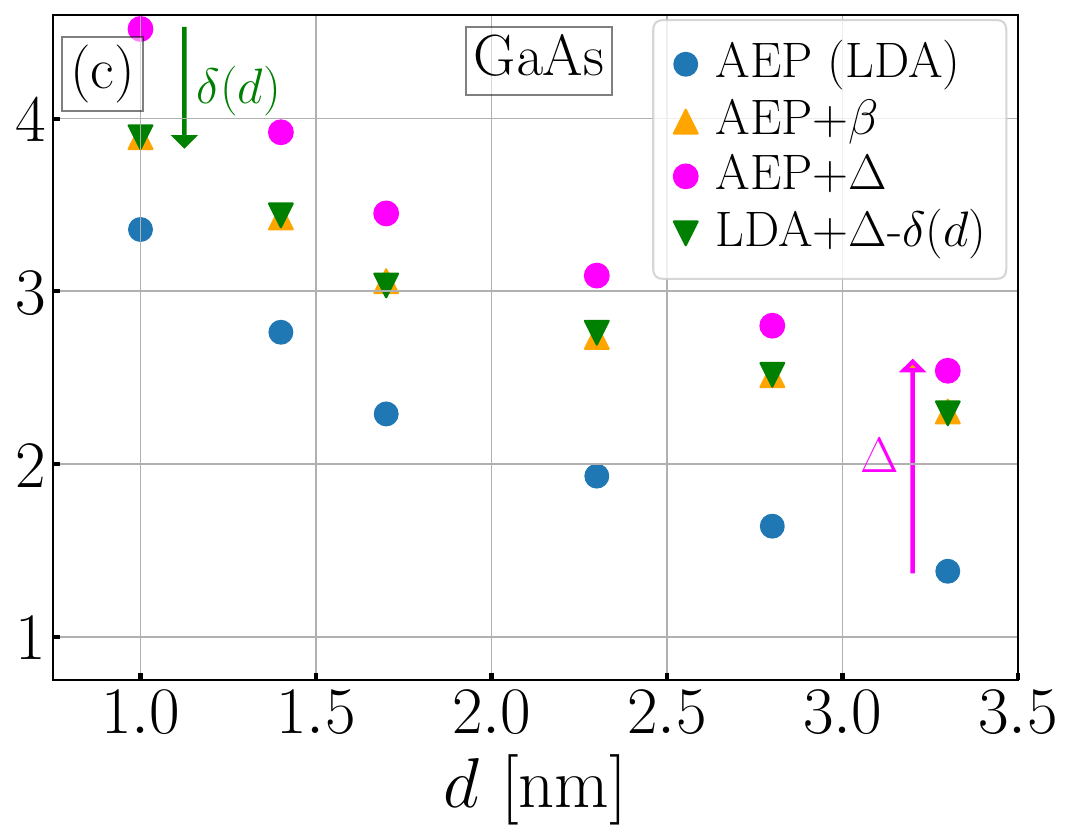}
\caption{ Single-particle band gap calculated with AEP (LDA),  "$\beta$-correction" (AEP+$\beta$), using bulk scissor ($\Delta$) and with the empirical fit ($\delta(d)$) for (a) InP (b) CdSe and (c) GaAs QDs. }
\label{fig:all_delta}
\end{figure*}

Since we have accurate quasiparticle and optical band gaps, as well as the (inaccurate) LDA results, we proceed by generating a simple correction term that can be used to improve LDA results.  
In Fig.~\ref{fig:all_delta} we show the quasiparticle band gaps of InP, CdSe and GaAs QDs calculated using different approaches. The correct gaps are given by the AEP $+ \beta$ results (yellow triangles). The AEP (LDA) results (blue solid circles) significantly underestimate the gap, as commonly known. For the scissor $\Delta$ correction (magenta) we have added the bulk band gap LDA error of 1.16 eV for GaAs, 1.15 eV for InP, and 1.70 eV for CdSe to the calculated QD band gaps. 
This procedure leads to an overestimated band gap, especially for smaller QDs, which can be understood from the too-low effective masses at the LDA level. The latter leads to an overestimation of the confinement effect and consequently too large band gap. 

To improve the LDA results, we suggest starting from the scissor-corrected LDA results and adding a size-dependent correction $\delta(d)$:
\begin{equation}
    E^\textrm{exact}_\textrm{QP gap} =E^{\rm LDA}_{ \rm gap} (d) +\Delta-\delta(d), \hspace{0.2 cm} \, \, \delta(d)=\frac{\rm A}{ d^{\rm x}} ,
    \label{eq:fit_eq}
\end{equation}
where $E^\textrm{exact}_\textrm{QP gap}$ represents the exact quasiparticle band gap and $E^{\rm LDA}_{\rm gap}$ represents the single-particle band gap calculated using LDA for a QD with diameter 
$d$ given in nm. We have utilized our AEP (LDA) and AEP+$\beta$ results to fit $\delta(d)$ and give the parameters in
 Table~\ref{tab:delta_fit}. The empirical fit (green triangles in Fig.~\ref{fig:all_delta}) and the exact results (yellow triangles) are in very good agreement for all the structures and materials. 

\begin{table}[h]
\centering
\begin{tabular}{ccccc}
 Materials  & A & x   & B& y\\
 \hline
 InP &0.656 & 0.609 &0.577 & 1.181\\
  \hline
 CdSe & 0.692   & 0.608 &0.846 &1.185\\
\hline
 GaAs& 0.627   & 0.767 &0.443 &1.139\\
\end{tabular}
\caption{ Fitting parameters used for the empirical quasiparticle band gap correction 
    (Eq.~\ref{eq:fit_eq}) and for the optical band gap correction (Eq.~\ref{eq:optical_corretion}) for InP, CdSe and GaAs. }
    \label{tab:delta_fit}
\end{table}
\begin{figure*}[ht]
\centering
\includegraphics[width=0.32\linewidth]{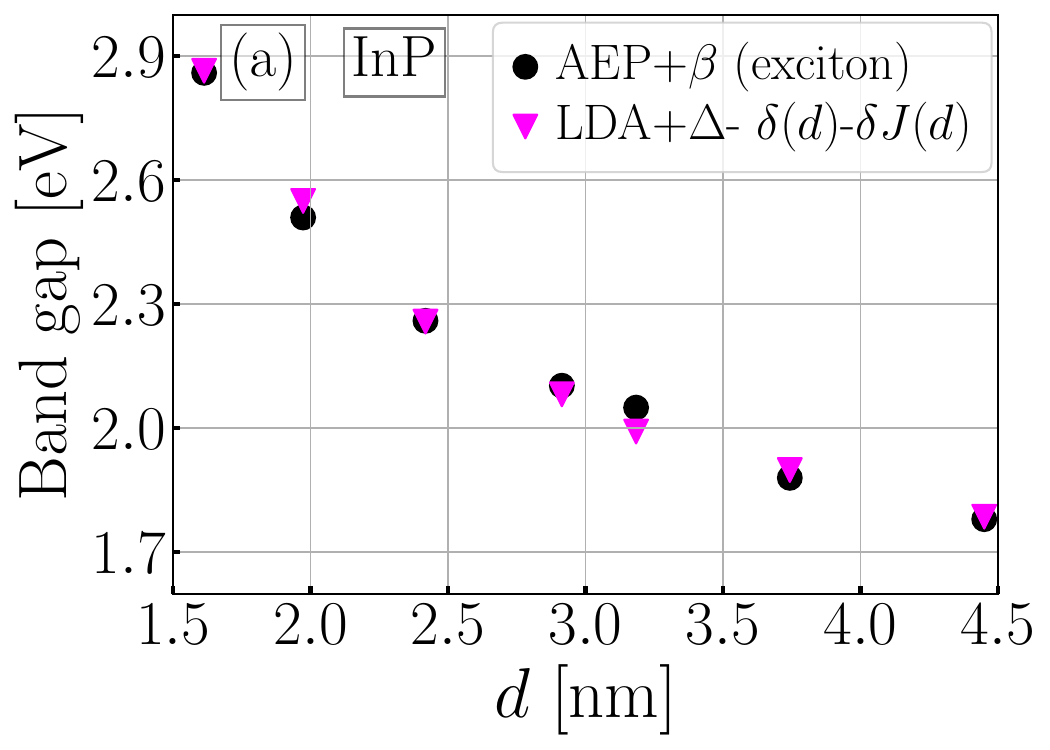}
\includegraphics[width=0.32\linewidth]{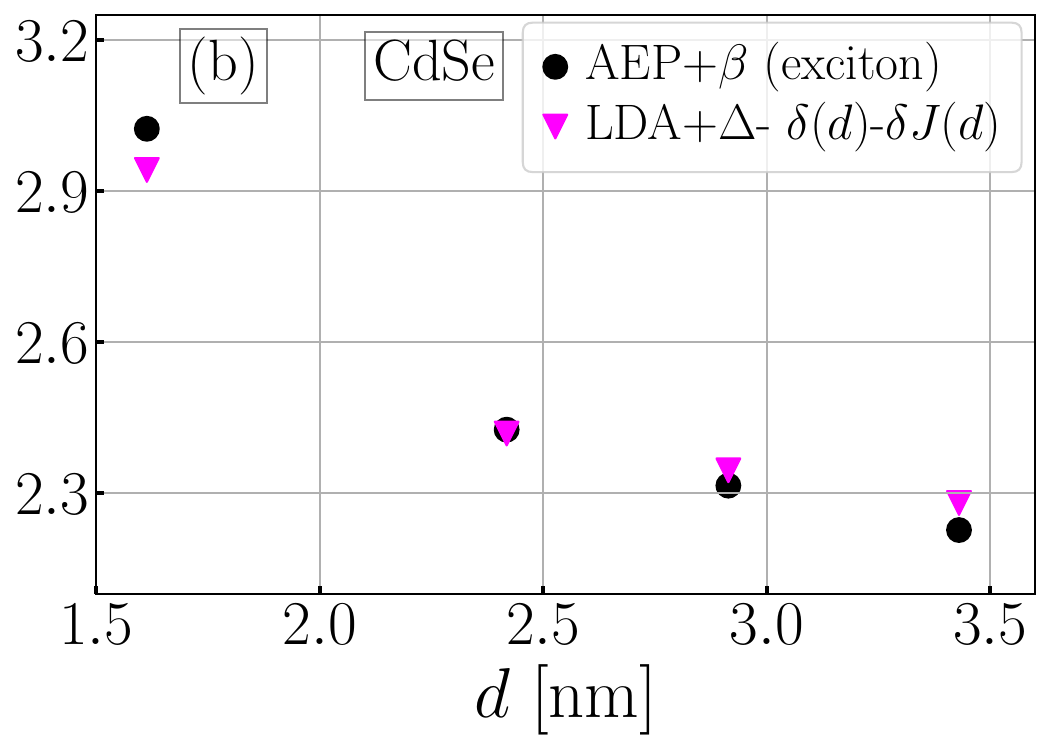}
\includegraphics[width=0.32\linewidth]{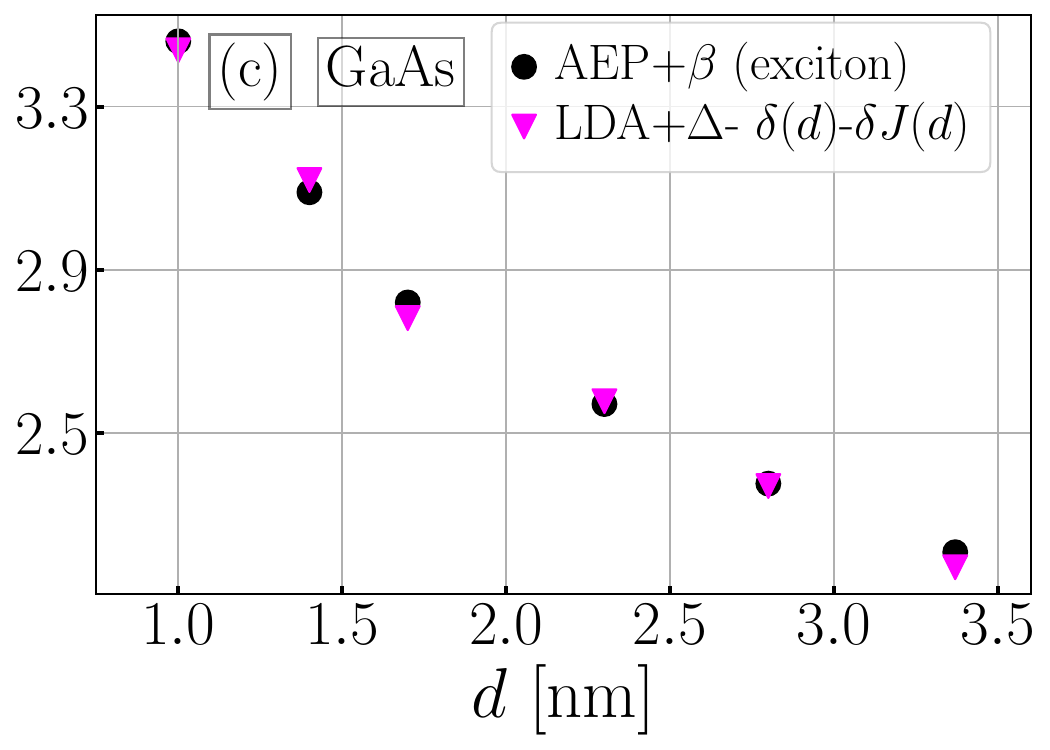}
\caption{Comparison of optical band gap calculated with our $\beta$-correction, obtained with fitted results from Eq.~(\ref{eq:optical_corretion})  for (a) InP (b) CdSe and (c) GaAs QDs.   }
\label{fig:OpticalGaps}
\end{figure*}
\subsection{Optical Band Gap}

While the quasiparticle band gap is relevant for electron affinities, work functions, and charging effect, the optical properties require to take excitonic effects into account. By using our accurate optical band gaps, calculated at the screened CI level, we can derive an empirical correction to the quasiparticle band gap obtained in the previous section. We use the simple fitting function:
\begin{equation} 
E^{ \rm optical}_{\rm gap }(d) = E^\textrm{exact}_\textrm{QP gap}  - \delta J(d), \qquad  
\delta J(d)=\frac{ \rm B}{d^{\rm y}},
    \label{eq:optical_corretion}
\end{equation}
where $B$ and $y$ are fitting parameters reported in Tab.~\ref{tab:delta_fit}.  In Fig.~\ref{fig:OpticalGaps} we show the QD optical band gap calculated using our correction and obtained from Eq.~\ref{eq:optical_corretion}. The exact results (black circles) and the result of the empirical fit (magenta triangles) are generally in very good agreement. 
In the supporting Information~\cite{noteSI} we show the raw data used to fit $\delta(d)$ and $\delta J(d)$.

\section{Conclusion}

We propose a correction scheme that can improve key properties of LDA-derived AEPs, 
including band gaps at different high-symmetry points of the Brillouin zone and effective masses. The approach is based on a modification of the non-local part of the pseudopotential and is restricted to the atomic core region. We demonstrate the accuracy of the methods by direct comparison to experiment for optical band gaps, intraband ($e_0$-$e_1$) splitting, Coulomb integrals, and excitonic fine structure of QDs with 1.5 to 4.5 nm diameter. Furthermore, a straightforward analytic expression to determine accurate quasiparticle and optical band gaps for InP, CdSe, and GaAs QDs from standard LDA calculation is provided. 

\section*{Acknowledgement}
Most computations were performed on the HPC cluster of the Regional Computing Center of the Universit\"at Hamburg. This work is supported by the Cluster of Excellence “Advanced Imaging of Matter” of the Deutsche Forschungsgemeinschaft (DFG) - EXC 2056–Project 390715994 and the DFG project GZ: BE 4292/4-1 AOBJ: 651735 ``Resonant Raman spectroscopy as a tool to investigate colloidal semiconductor nanocrystals".

\section*{CRediT authorship contribution statement}
\noindent
\textbf{Surender Kumar:} Conceptualization, Investigation, Formal analysis, Writing - Original Draft. \textbf{Hanh Bui: }Conceptualization, Formal analysis, Writing – review and editing,   \textbf{Gabriel Bester: } Conceptualization, Supervision, Resources, Validation, Writing – review and editing. 
 
\bibliographystyle{elsarticle-num} 
\bibliography{Gap_correction}


\end{document}


\begin{frontmatter}

\title{Supporting Information: Empirical Band-Gap Correction for LDA-Derived Atomic  Effective Pseudopotentials}

\author[inst1]{Surender Kumar}

\affiliation[inst1]{organization={Departments of Chemistry and Physics, Universität Hamburg},
            addressline={Luruper Chaussee 149}, 
            city={Hamburg},
            postcode={D-22761}, 
            country={Germany}}

\author[inst1]{Hanh Bui}

\author[inst1,inst2]{Gabriel Bester \corref{c1}}
\cortext[c1]{Corresponding author}
\ead{gabriel.bester@uni-hamburg.de}

\affiliation[inst2]{organization={The Hamburg Centre for Ultrafast Imaging},
            addressline={Luruper Chaussee 149}, 
            city={Hamburg},
            postcode={D-22761}, 
            country={Germany}}

\end{frontmatter}

\section{Effective Mass calculations}

To compare with our atomistic results, we have used infinite spherical well (ISW) and finite spherical well (FSW) (with a height of 0.7 eV) effective mass approximations to calculate the band gap for InP quantum dots (QDs). For the ISW case, we have used  the following equation:
\begin{align}
    E_{nl}=\frac{\hbar^2 }{2m^*_{e/h}\,a^2} z^2_{nl}  
\end{align}
where {\it{z$_{nl}$}}  is the n-th root of the Bessel  Function, $a$ is the radius of the nanocrystal and {\it{m$^*_{e/h}$}} is the effective mass of the electron/hole.

For FSW, we use  the following equations:
\begin{subequations}
\begin{align} \label{eq:ISW1}
-k \cot(ka) & = q \hspace{2.4cm}\textrm{for } l=0 \textrm{ case} \\
k^{-2}(1-k a \cot(ka)) & = -q^{-2}(1+qa) \hspace{0.5cm}\textrm{for } l=1 \textrm{ case} \label{eq:ISW2}
\end{align}
\end{subequations}
where
\begin{align*} 
 k^{2}=\frac{2 m^*_{e/h} }{\hbar^{2}} (E+V_0)\quad 
 \end{align*}
 \begin{align*}
 q^{2}=-\frac{2 m^*_{e/h} }{\hbar^{2}}E \quad ,
\end{align*}
where V$_0$ is the height of the confining barrier. Similar to FSW, the solution of ISW cannot be obtained analytically. Instead, one needs to solve Eq.~(\ref{eq:ISW1}) numerically.

\newpage
\section{Wavefunction overlap before and after $\beta$-correction}

In table~\ref{tab:overlap} we show the overlap $\langle\Psi_{\rm AEP}|\Psi_{\rm  AEP+\beta} \rangle$ ( in $\%$ ) for $e_0$ and $h_0$ in InP quantum dots of different sizes, along with the bulk wavefunction overlap at the $\Gamma$-point. We see no trend with the QD size and obtain a value between 98.0 \% and 99.3 \%.
\begin{table}[!hb]

    \centering
    \begin{tabular}{p{2.75cm} |p{1.5cm}   |p{1.5cm} }
    \hline
    \hline
      Diameter (nm) & $e_0$  & $h_0$  \\
      \hline
      \hline
      bulk & 99.8 & 99.8   \\
      \hline
      1.61 & 99.3&  99.1 \\
      \hline
      1.97 & 98.8  & 98.7  \\
      \hline
       2.42 & 99.3  & 98.7 \\
      \hline
      2.91 & 98.0 & 98.3 \\
      \hline
     3.18 &  98.3 & 98.3  \\
      \hline
      3.74 & 99.1 & 98.1 \\
     \hline
      \hline
    \end{tabular}
    \caption{Wavefunction overlap  $\langle\Psi_{\rm AEP}|\Psi_{\rm  AEP+\beta} \rangle$ ( in $\%$ ) for $e_0$ and $h_0$ states before and after the correction for InP bulk and QDs. }
    \label{tab:overlap}
\end{table}

\section{Band gap  and Coulomb integral corrections} 

\begin{figure}[h]
\centering
\includegraphics[width=0.445\linewidth]{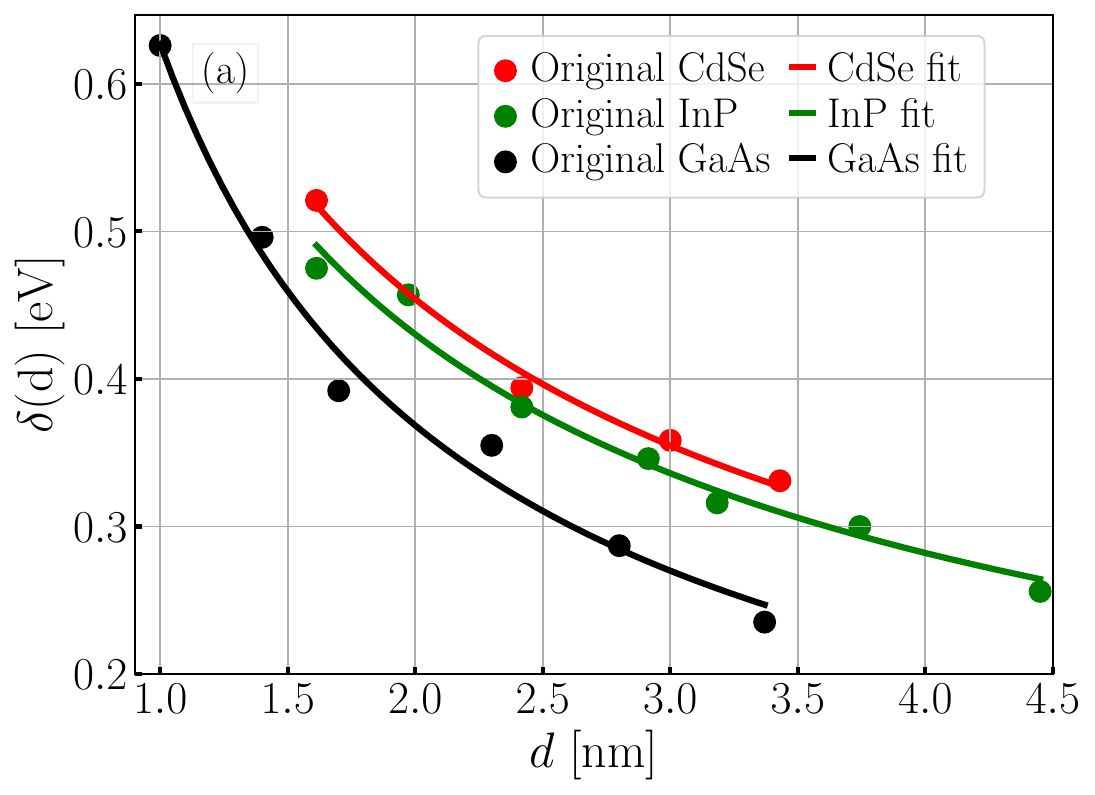}
\includegraphics[width=0.450\linewidth]{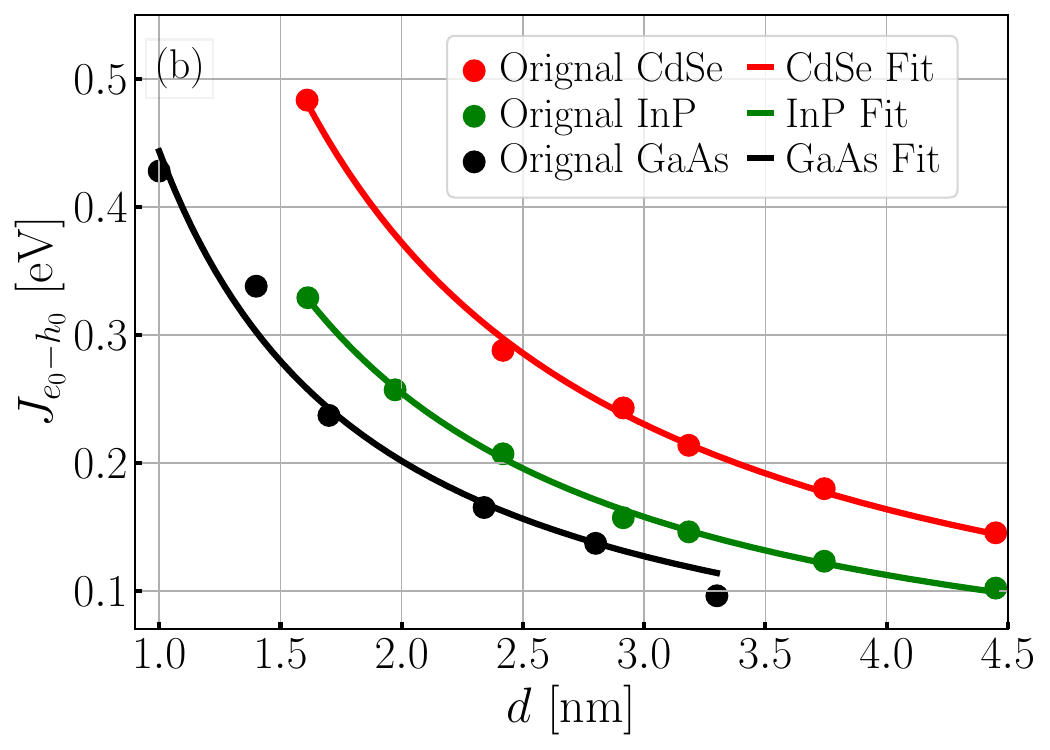}
\caption{Fitted plot for (a) the empirical band gap fit 
    $\delta(d) = \text{A}/d^x$ (Eq.~(7) in the main text)  (b)  the empirical Coulomb integrals fit $\delta J(d) = \text{B}/d^y$  (Eq.~(8) in the main text) for InP, CdSe and GaAs.}
\label{fig:all_delta}
\end{figure}

In Fig.~\ref{fig:all_delta} we plot the fitting curves $\delta(d) = \text{A}/d^x$ and  $\delta J(d) = \text{B}/d^y$ used to improve the quasiparticle and the optical gaps in the main manuscript. The data points are the numerical atomistic results and are shown to lie very close to the curves, except for GaAs that show some deviations. These deviations are possible, since the diameter dependence in an atomistic description involves the abrupt addition of atomic shells and possible sudden changes of composition. 

\bibliographystyle{unsrt}


